# Generative Large Language Models for Knowledge Representation: A Systematic Review of Concept Map Generation

Xiaoming Zhai, AI4STEM Education Center, University of Georgia, Athens, 30602

*Abstract*

The rise of generative large language models (LLMs) has opened new opportunities for automating knowledge representation through concept maps, a long-standing pedagogical tool valued for fostering meaningful learning and higher-order thinking. Traditional construction of concept maps is labor-intensive, requiring significant expertise and time, limiting their scalability in education. This review systematically synthesizes the emerging body of research on LLM-enabled concept map generation, focusing on two guiding questions: (a) What methods and technical features of LLMs are employed to construct concept maps? (b) What empirical evidence exists to validate their educational utility? Through a comprehensive search across major databases and AI-in-education conference proceedings, 28 studies meeting rigorous inclusion criteria were analyzed using thematic synthesis. Findings reveal six major methodological categories: human-in-the-loop systems, weakly supervised learning models, fine-tuned domain-specific LLMs, pre-trained LLMs with prompt engineering, hybrid systems integrating knowledge bases, and modular frameworks combining symbolic and statistical tools. Validation strategies ranged from quantitative metrics (precision, recall, F1-score, semantic similarity) to qualitative evaluations (expert review, learner feedback). Results indicate LLM-generated maps hold promise for scalable, adaptive, and pedagogically relevant knowledge visualization, though challenges remain regarding validity, interpretability, multilingual adaptability, and classroom integration. Future research should prioritize interdisciplinary co-design, empirical classroom trials, and alignment with instructional practices to realize their full educational potential.

**Introduction**

In recent years, the advancement of generative large language models (LLMs), such as GPT-4, has reshaped the educational landscape, especially in tasks involving knowledge organization, representation, and visualization (Lee et al., 2025; Rozić et al., 2023). One particularly promising application is in the construction of concept maps—graphical tools widely used in educational settings to facilitate meaningful learning, illustrate relationships among key concepts, and foster higher-order thinking skills (Novak & Gowin, 1984; Wang et al., 2025; Yang et al., 2025). However, despite their pedagogical value, the creation of concept maps typically requires substantial cognitive engagement, deep domain understanding, and time investment, which have posed challenges to their routine use by both educators and learners (Schroeder et al., 2018). Generative LLMs provide a compelling solution to these limitations by automating the extraction and structuring of conceptual knowledge from unstructured texts (Perin et al., 2023). This capability has the potential to democratize the use of concept mapping across diverse contexts—enabling personalized, scalable, and cost-effective knowledge representations that were previously infeasible in traditional classroom environments.

LLMs leverage pre-trained knowledge on vast and diverse corpora, which enables them to capture semantic connections across domains and articulate them in ways that align with human reasoning (OpenAI, 2023). These models can infer implicit and explicit relationships among concepts, articulate them in semantically coherent ways, and generate visualizable outputs that mirror human-authored maps in clarity and depth. Additionally, LLMs afford scalability for real-time, low-cost generation of concept maps tailored to various instructional goals, thereby reducing barriers to implementation (Galletti et al., 2025). The generative nature of LLMs also allows for dynamic refinement of outputs, which is particularly useful for iterative learning and formative assessment. When integrated into educational workflows, LLM-generated concept maps can serve as both diagnostic tools for instructors and scaffolding mechanisms for learners (Faraji et al., 2025; Ferentinou et al., 2025).

While preliminary studies have explored the usability of LLMs for concept mapping (Bayrak & Dal, 2025), the existing literature is fragmented and lacks a comprehensive synthesis of the emergent technical methods and validation strategies. Most current work focuses on isolated use cases or exploratory applications, without systematically comparison of the methodological approaches or examining the underlying mechanisms that drive the performance of LLM-generated concept maps (Chang et al., 2025; Chiou & Lin, 2025). Moreover, empirical evaluations of LLM-generated maps are still limited, and key questions remain about their validity, reliability, and alignment with pedagogical standards. This fragmented understanding reveals two critical areas where further research is needed: First, there is a lack of systematic review of the diverse techniques and LLMs used to generate concept maps with LLMs, including how these approaches differ in terms of structure, complexity, and alignment with human-generated standards. Second, research is needed to understand the current methods to validate LLM-generated concept maps. Addressing these issues is vital to advancing both the theoretical and practical knowledge base surrounding the use of LLMs in educational settings.

This review study addresses these gaps by answering two research questions: (a) What emergent methods of LLMs, including their technical features, have been employed to construct concept maps? (b) What empirical evidence has been drawn to validate the LLM-generated concept maps? What are the pros and cons of LLM-generated concept maps compared to those constructed using traditional methods? Through a systematic synthesis of existing studies, this review aims to illuminate the current state of research and provide directions for future investigations on the integration of LLMs in concept mapping for educational and cognitive purposes.

**Concept Maps**

Concept map is a structured representation of knowledge with visually displayed relationships (links) among concepts (nodes) originally proposed by Joseph D. Novak and colleagues (Novak & Gowin, 1984), grounded in Ausubel's theory of meaningful learning (Ausubel, 1963). Concept map is both a learning tool and a knowledge representation method, which helps learners externalize and organize their understanding, supporting deeper learning by explicitly mapping out how ideas are connected (Chang et al., 2023; Chu et al., 2025; Su & Zou, 2024). For example, Ma et al. (2025) developed an AR-based learning tool, PeachBlossom, that incorporated concept maps to support students' scientific concept learning. After testing its effectiveness in a quasi-experiment with 85 seventh graders, they found that the concept map strategy benefited students with low to medium prior mental models and reduced their mental effort, though it did not significantly affect mental load.

According to Novak and Cañas (2008), the common procedure for constructing a concept map involves five steps: identifying a specific knowledge domain, compiling key concepts, organizing these concepts hierarchically, adding linking phrases to represent relationships, and revising the map to reflect additional insights. While this process is pedagogically valuable—fostering deeper understanding and enabling students to externalize and reflect upon their cognitive structures—it is also labor-intensive and requires significant domain expertise. This has limited its scalability and practicality, especially when applied to large datasets or integrated into routine classroom instruction (Chu et al., 2025). As educational environments increasingly incorporate technology to enhance learning efficiency and accessibility, the need for more scalable and automated approaches to concept map construction becomes evident.

To meet this growing demand for scalable and automated solutions in education, Natural Language Processing (NLP) has emerged as a foundational technique for transforming unstructured textual data into structured conceptual representations suitable for concept map generation (Kornilakis et al., 2004). Drawing from linguistic theory, NLP encompasses a suite of analytical tools—including part-of-speech tagging, syntactic parsing, dependency analysis, and semantic role labeling—that allow for deep analysis of linguistic structures (Dos Santos et al., 2019). These tools help identify key concepts within text, clarify their grammatical roles, and reveal relationships that mirror how humans make sense of content. Beyond extraction, NLP also supports the organization of concepts into hierarchical structures, which is central to effective concept mapping. The ability to interpret language with such nuance is particularly valuable in educational contexts where precision, clarity, and contextual accuracy are essential (Nugumanova et al., 2021; Wang et al., 2008).

In tandem with NLP, data-driven methods like text mining have become increasingly influential in knowledge representation analysis, particularly in concept map generation. Text mining contributes a complementary dimension by leveraging statistical and machine learning techniques to discover hidden patterns, topic clusters, and co-occurrence relationships that may not be readily identifiable through linguistic parsing alone (Samrat, 2023; Wang et al., 2008). Common techniques such as term frequency-inverse document frequency (TF-IDF), latent semantic analysis (LSA), and topic modeling help uncover which terms are most salient and how often they appear together in contextually significant ways. These methods are particularly advantageous when analyzing extensive textual repositories—such as student essays, digital textbooks, or discussion forums—where patterns of meaning emerge from aggregation rather than rule-based interpretation (Kasinathan et al., 2022). When integrated with NLP, text mining strengthens the reliability and validity of concept extraction, enabling more comprehensive, scalable, and contextually relevant concept map construction. The hybridization of these approaches aligns well with the growing emphasis on evidence-based instructional design and opens new pathways for leveraging educational texts as data-rich sources of cognitive insight.

Building on these foundations, generative AI significantly extends the potential of automated concept mapping. Generative AI models, particularly those leveraging (NLU) and Natural Language Generation (NLG), offer an end-to-end framework for interpreting, synthesizing, and articulating knowledge from educational content. NLU focuses on decoding grammatical and semantic structures, enabling machines to grasp meaning and resolve ambiguities in context-rich environments (Weld et al., 2022). Meanwhile, NLG translates structured data into coherent, fluent, and pedagogically appropriate text (Dong et al., 2022). Applied to concept mapping, these capabilities allow AI systems to autonomously extract conceptual structures and articulate them in ways that support learning and instruction. The generative affordances also introduce dynamic adaptability, enabling the creation of tailored concept maps that align with diverse instructional goals and learner profiles.

Supporting this technological evolution, Dos Santos et al. (2019) conducted a systematic review of 23 studies applying NLP to concept map development. They proposed a classification framework that distinguishes between tools focused on concept identification and those aimed at extracting inter-concept relationships, further categorizing them by their reliance on linguistic versus statistical methods. Linguistic approaches draw on grammatical and syntactic rules, while statistical methods infer relationships based on frequency and co-occurrence data. This classification underscores the methodological richness of the field and highlights the potential benefits of hybrid models that combine the rule-based accuracy of linguistic methods with the scalability of statistical algorithms. When enhanced by the adaptive capacities of LLMs, such hybrid approaches can substantially improve the quality, interpretability, and pedagogical value of automatically generated concept maps in educational settings. Thus, the integration of NLP, text mining, and generative AI technologies presents a comprehensive, scalable, and pedagogically sound pathway for the automated construction of high-quality concept maps.

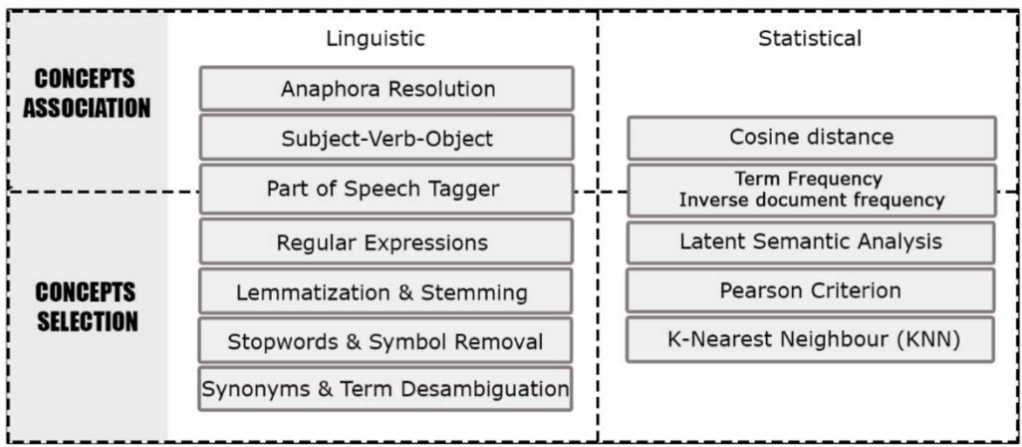

*Figure 0-1 A classification schema for concept map generation using NLP (Dos Santos et al., 2019)*

Nugumanova et al. (2021) reviewed relevant studies between 2001 and 2020 and found approximately 50 works were published on the topic of automatic construction of concept maps from natural language texts. They found that most of the identified publications utilized morphological and syntactic analysis to extract concepts and relationships from texts (e.g., Aguiar et al., 2016). Common techniques included part-of-speech (POS) tagging, syntax tree construction, and extraction of morpho-syntactic patterns. These methods were often complemented by additional approaches such as co-reference resolution, synonym extraction, and named entity recognition (e.g., Falke & Gurevych, 2017). Some studies also employed association rule mining to infer relationships between terms (e.g., Shao et al., 2020) or leveraging rule-based data augmentation approach (Shao et al., 2024).

The extracted concepts and relationships were frequently grouped into categories and ranked by importance. Clustering techniques were commonly used for grouping (Pipitone et al., 2014), while statistical methods like TF-IDF (Falke, 2019), LSA (Pipitone et al., 2014), PCA (Chen et al., 2008), and burst analysis (Nugumanova et al., 2021) were employed for ranking. Ultimately, these concepts and relationships were integrated into concept maps, often modeled as graphs, to visually represent the extracted knowledge.

Despite the promises of the automatic concept map generations, traditional approaches often rely on auxiliary resources and carefully crafted heuristics (Falke, 2019; Falke & Gurevych, 2017; Lu et al., 2023; Pipitone et al., 2014). The separation of concept map construction from downstream tasks can cause discrepancies between the generated maps and the task requirements. Additionally, traditional methods often generate redundant or verbose nodes due to their dependence on frequency-based or ad hoc pipelines. Recent research has focused on automatically generating concept maps using weak supervision from downstream tasks. One such approach, Doc2Graph (Gemelli et al., 2022; Yang et al., 2020), employs an end-to-end neural network model. While this method enables scalable concept map generation, it often suffers from semantic incompleteness and noisy connections due to a lack of linguistic analysis. Furthermore, Doc2Graph is not label-efficient, requiring significant weak supervision for meaningful concept maps. Its fixed-size graph design further limits flexibility, whereas the ideal graph size should adapt to the complexity of the underlying documents.

**Generative Large Language Models for Automatic Generation of Concept Maps**

Building upon the limitations of traditional concept mapping methods and the evolving landscape of NLP and text mining techniques, generative LLMs offer a significant leap in automating concept map construction. Unlike conventional approaches that often rely on rule-based heuristics, external ontologies, or weak supervision from downstream tasks (Yang et al., 2020), LLMs possess the unique ability to generate semantically coherent and contextually meaningful conceptual structures directly from unstructured text. This generative capacity is enabled by their pretraining on vast, diverse corpora and their transformer-based architectures, which model deep semantic relationships and long-range dependencies in language (Brown et al., 2020). As such, LLMs are not merely extractive systems but function as sophisticated generative engines capable of synthesizing new knowledge structures. They are particularly adept at discerning nuanced semantic patterns, enabling the distillation of complex ideas into concise conceptual representations. This ability is central to effective concept map generation, especially when aiming to reflect higher-order thinking or disciplinary discourse.

  What further distinguishes LLMs from previous automatic concept mapping tools is their capacity to streamline the entire pipeline—from concept identification and relationship extraction to hierarchical structuring—within a unified, generative framework(Cheng et al., 2024). This integration reduces the need for handcrafted rules and modular dependencies, allowing for greater fluidity, adaptability, and scalability. Additionally, LLMs can dynamically tailor the content, scope, and complexity of generated maps based on domain-specific prompts, user needs, or pedagogical goals (Ma & Chen, 2025). This contextual sensitivity is particularly advantageous for educational applications, where instructional alignment, differentiated instruction, and learner diversity are critical. For example, an LLM could generate one version of a concept map to support novice learners and a more complex version for advanced learners—customized not only in content but also in structure and explanatory depth.

Moreover, LLMs have demonstrated strong zero-shot and few-shot learning capabilities, enabling them to construct relevant concept maps even with minimal task-specific data or fine-tuning (Galletti et al., 2025). These models excel at extrapolating patterns from limited input and aligning their outputs with inferred instructional goals (Schicchi et al., 2025). Such capabilities support real-time feedback, formative assessment, and adaptive learning environments—critical features of modern, technology-enhanced learning ecosystems. For instance, LLM-generated maps can scaffold students' understanding by making explicit the relationships among implicitly linked concepts, enhancing metacognitive awareness. They can also assist teachers in diagnosing misconceptions through automatically generated visual summaries, thus enabling timely pedagogical interventions.

Despite these promising affordances, current implementations remain challenged by issues related to explainability, consistency across outputs, and generalization to specific domains or curricula. Current studies are largely exploratory and vary in terms of evaluation metrics, model configurations, theoretical foundations, and use contexts (Nugumanova & Baiburin, 2021). Few empirical studies have rigorously evaluated the pedagogical utility or learning outcomes associated with LLM-generated maps. A systematic synthesis is urgently needed to consolidate findings, categorize methodologies, and assess the educational value of LLM-generated maps. This review addresses that gap by critically examining the technical strategies employed in current research, identifying the strengths and limitations of LLMs in this space, and evaluating the extent to which these systems meet the pedagogical criteria for effective knowledge representation. By doing so, this review aims to inform the development of more interpretable, scalable, and pedagogically sound approaches for leveraging generative AI in educational settings, ultimately supporting future research and implementation in real-world classrooms.

**Methods**

**Literature Collection Criteria**

*Inclusion Criteria*

To ensure a focused and rigorous review of relevant studies, the inclusion criteria were defined as follows:

- Studies that explicitly employed generative LLMs, such as GPT-series, BERT, or similar transformer-based architectures, in the construction of concept maps.
- Studies that provided sufficient technical descriptions of how LLMs were utilized in generating concept maps, including model type, input format, mapping algorithms, evaluation metrics, or integration strategies.
- Articles for which the full text was accessible, either through institutional access or open-source repositories, to allow for complete methodological and analytical examination.
- Papers written in English to maintain consistency and interpretability of findings across reviewed literature.

*Exclusion Criteria*

The following types of studies were excluded from the review:

- Papers that mentioned LLMs but did not involve their use in generating concept maps or only applied LLMs to unrelated tasks such as summarization or classification.
- Studies that lacked technical clarity or depth regarding the operationalization of LLMs in the concept map generation process, such as vague descriptions or speculative applications.

- Review articles, theoretical discussions, or conceptual papers that did not present original empirical findings or implementation results related to the generation of concept maps using LLMs.

These criteria were developed to ensure the inclusion of studies that not only addressed the core topic of LLM-enabled concept map generation but also contributed substantive insights into the technical and educational implications of this emerging practice.

*Literature Search and Selection*

The literature search process followed best practices for systematic reviews in the education research field, with an emphasis on transparency, replicability, and comprehensiveness. Searches were conducted in multiple scholarly databases commonly used in educational technology and learning sciences, including Web of Science, Scopus, ERIC, and Google Scholar. To ensure coverage of both peer-reviewed journal articles and conference proceedings, the search also included proceedings from major AI education conferences such as AIED, LAK, and EDM. Keywords used in the search strategy included combinations of terms such as "*concept map*," "*knowledge map*," "*generative language model*," "*large language model*," "*GPT*," "*transformer*," "*automatic generation*," and "*education*." Boolean operators (e.g., AND, OR) and truncation symbols were applied to capture a broad set of relevant results. For example, a representative query was: ("*concept map*" OR "*knowledge map*") AND ("*large language model*" OR "*LLM*" OR "*GPT*") AND ("*education*" OR "*instruction*"). The search was restricted to studies published between 2018 and 2025 to reflect the emergence and evolution of transformer-based LLMs in education research.

The initial search yielded 214 unique records from the selected databases. After removing 43 duplicate entries, 171 records remained for initial screening. Titles and abstracts of these articles were independently reviewed by two researchers to assess their relevance to the scope of the study. Disagreements were resolved through discussion and, where necessary, consultation with a third reviewer. This preliminary screening process led to the exclusion of 87 articles that either did not involve the use of generative LLMs for concept mapping or failed to meet other inclusion criteria, resulting in 84 articles retained for full-text analysis. Each of these 84 full-text articles was carefully evaluated against the inclusion and exclusion criteria. The evaluation process focused on examining the technical depth, model description, input-output procedures, evaluation metrics, and the educational context in which the LLMs were deployed. Studies lacking clear methodological transparency or empirical relevance were excluded, leading to the final selection of 28 studies that provided substantive contributions to the research questions. To enhance the completeness of the review, backward reference searches were also conducted on all included articles, identifying two additional relevant papers that were subsequently added to the final pool.

All stages of the literature search and selection were documented following the PRISMA 2020 guidelines. A flow diagram was created to depict the systematic filtering of studies across four phases: identification, screening, eligibility, and inclusion. This methodological transparency not only increases the reproducibility of the study but also ensures the integrity and reliability of the review process. By adhering to these structured procedures, the study presents a comprehensive, methodologically sound, and unbiased synthesis of current research on LLM-based concept map generation in educational contexts.

*Data Analysis*

To address the two guiding research questions, this review employed a thematic synthesis methodology, widely used in qualitative educational research to systematically integrate findings across diverse empirical studies (Thomas & Harden, 2008). This method was selected for its ability to extract meaningful patterns, techniques, and interpretations from a heterogeneous body of literature, which is particularly relevant given the varied use cases, models, and evaluation frameworks present in studies employing LLMs for concept map generation.

To investigate the first research question—concerning the methods and technical attributes of LLMs used for concept map generation—an inductive coding strategy was applied. Studies were analyzed for key methodological details such as model architecture (e.g., GPT, BERT), training and fine-tuning processes, prompt design, input-output format specifications, and the overall pipeline structure (e.g., multi-step versus end-to-end). This allowed for the emergence of categories based on how LLMs are technically operationalized across studies. Thematic analysis helped capture the diversity and innovation in design practices, including prompt engineering, adaptive generation workflows, and integration with educational platforms.

To examine the second research question—focused on the empirical validation, benefits, and drawbacks of LLM-generated concept maps—studies were analyzed for their evaluation frameworks and validation strategies. Coding categories included the type of evaluation used (e.g., expert review, learner feedback, rubric-based scoring), the comparative analysis between human- and machine-generated maps, and the reported alignment with educational goals such as scaffolding, assessment, or engagement. This stage of the analysis also considered the presence of challenges like semantic drift, hallucinated nodes, or structural inconsistency.

To ensure analytic rigor, two researchers independently reviewed and coded the included articles. Coding reliability was strengthened through iterative comparison, discussion, and refinement of categories. The final thematic framework was shaped by consensus and supported by examples drawn from across the dataset. This procedure enabled the review to systematically respond to the research questions while maintaining methodological transparency, rigor, and relevance to educational research.

**Findings**

*Methods and technical attributes of LLMs in construction concept maps*

    *Human-in-the-Loop*. The AIMindmaps (Lin et al., 2022) system is designed to automatically generate mind maps from user-uploaded PDF documents using LLMs and a generative model. Users can customize the process by configuring preferences, such as the number of mind map layers, and further refine the generated mind maps to meet their needs. The system comprises a front-end user interface and a back-end server for processing PDF documents and generating mind maps. To use the system, users upload a PDF document and specify their desired configurations through the configuration panel. These configurations are sent to the back end, which guides the generation of the mind map. This user-centric approach allows individuals to actively participate in the creation process by setting their preferences within a flexible framework.

    The generated mind map includes a central image and associated text content. Users can refine the mind map using a tool panel that offers various editing features, such as modifying subtopics, adjusting relationships, creating or modifying images, and adding notes. For instance, users can use the image generation panel to customize the central image or add AI-generated symbols to other subtopics, addressing challenges users might face due to limited artistic capabilities or resources (Erdem, 2017). The option to add notes enhances the usability of the mind map by mitigating issues like forgetting the meaning of symbols. The system operates in four main steps: a) It reads the uploaded PDF document and extracts relevant information. b) A LLM is called using a structured prompt to identify and organize key details from the document.

c) The generative model combines these structured prompts and the main topic to create the central image. d) Finally, the processed information is sent to the front end for rendering, producing the completed mind map. This workflow ensures a seamless integration of user inputs, AI-powered content extraction, and visual generation, offering an intuitive tool for creating personalized and visually appealing mind maps.

*Weak-Supervised Learning*. The GT-D2G approach, short for Graph Translation-based Document-to-Graph, is a framework designed to generate high-quality concept maps from text documents under weak supervision from downstream tasks (Lu et al., 2023). Its purpose is to overcome the limitations of traditional and existing neural methods in concept map generation by producing task-oriented, semantic-rich, concise, label-efficient, and size-flexible graphs. The core mechanism of GT-D2G integrates an NLP pipeline to construct semantic-rich initial graphs and a graph translation model to iteratively refine these graphs.

The process begins with the Initial Graph Constructor, which employs NLP tools like constituency parsing and coreference resolution to extract meaningful concepts (e.g., noun phrases and verb phrases) and generate initial undirected graphs. These graphs serve as a rich starting point, embedding linguistic knowledge while maintaining flexibility for refinement. Next, the Graph Translator applies a graph pointer network and an adjacency vector generator to iteratively select key nodes and establish links, ensuring that the final graph captures task-relevant information. The Graph Encoder and Graph Predictor play key roles in this process, with the former learning node embeddings using Graph Convolutional Networks (GCN) and the latter applying a Graph Isomorphism Network (GIN) to optimize the graph for specific downstream tasks, such as document classification. The entire system operates in a weakly supervised, end-to-end manner, leveraging task-specific labels as guidance for graph refinement.

A notable feature of GT-D2G is its size-flexible graph generation, achieved through penalties on graph length and coverage to balance conciseness and completeness. Additionally, the framework incorporates techniques like Gumbel-softmax sampling for differentiable and stochastic node selection, further enhancing the quality and adaptability of the generated graphs. In experiments, GT-D2G demonstrates superiority over traditional and neural concept map generation methods, particularly in semantic richness, task alignment, and scalability. Its design ensures efficient labeling, robust downstream performance, and flexible graph sizes tailored to document complexity.

***Fine-tuned Large Language Models on Domain-specific Datasets***. One prominent category involves the fine-tuning of LLMs on domain-specific datasets that pair input texts with expert-created concept maps (Perin et al., 2023). This approach begins by collecting large annotated corpora in which concepts and the relationships among them are explicitly labeled—often in the form of triples or semantic frames. These datasets serve as training input for LLMs, such as fine-tuned variants of BERT, T5, or GPT models. The training process involves learning to predict structured representations—typically in graph-based or JSON formats—from raw text inputs. For example, Shi et al. (2025) introduce a dual data mapping system that uses fine-tuned large language models to automate the mapping between digital twins (DTs), asset administration shells (AAS), and existing software models. It demonstrates that this approach improves interoperability in manufacturing by enabling accurate, efficient, and scalable integration of AAS instances with existing systems. Despite its improved performance, the technical demands of this approach include a higher degree of domain-specific annotation, GPU resources for fine-tuning, and evaluation against gold-standard concept maps, using metrics such as F1-score, edge accuracy, and semantic coherence.

The methodological framework consists of several key stages. Initially, a pre-trained LLM, such as GPT-4, is selected based on its demonstrated ability to process complex sentence structures, recognize named entities, and extract relationships. The model undergoes a fine-tuning process using a curated dataset comprising unstructured text and corresponding structured propositions. Through this fine-tuning, the LLM learns to extract task-specific patterns, particularly tuples that represent concepts and their relational connections. This customization enables the model to extract semantic-rich propositions directly from raw text, which subsequently form the foundation of the concept map. The resulting propositions are visualized as nodes (concepts) and edges (relationships), yielding an intuitive and structured representation of the input text.

The key features of this approach include its capacity to produce semantically rich and task-specific propositions, its scalability for processing extensive volumes of text, and its adaptability across diverse domains through fine-tuning. Empirical evaluations highlight the superior precision and recall achieved by the fine-tuned LLM compared to traditional methods, reflecting its ability to minimize semantic noise and enhance the relevance of extracted propositions. Additionally, the approach supports flexible generation of concept maps, with the size and complexity of the maps tailored to the requirements of the input data and specific tasks.

**Pre-trained LLMs without additional model fine-tuning**. This strategy takes advantage of the zero-shot or few-shot generalization capabilities of models like GPT-3.5 or GPT-4. Carefully constructed prompts instruct the model to identify salient concepts and their interconnections within a given passage. The process typically starts by feeding the LLM with a text passage and a task-specific prompt such as "List all key concepts and describe their relationships in the form of a concept map." Some approaches utilize few-shot prompting, where the model is shown examples of concept mapping before performing the target task. As demonstrated by Chen et al. (2024), prompt-based LLMs can be used to generate pedagogically useful concept structures from open-domain and educational texts, which support students' problem-solving in STEM education. Technically, this method is computationally efficient and requires no training data but is sensitive to the clarity and specificity of prompts. Tools like LLMapper exemplify this method, integrating GPT-generated outputs with interactive visualization platforms to allow users to inspect and refine the concept maps generated from encyclopedic or instructional content (ARANGO, 2024).

***LLMs with external structured knowledge sources.*** This approach leverages resources such as ontologies, taxonomies, or knowledge graphs, to enhance the quality and reliability of generated concept maps. In this hybrid methodology, the LLM serves as an initial parser that extracts candidate concepts and potential relationships from raw textual inputs. These extracted elements are subsequently validated, corrected, or supplemented through alignment with structured semantic databases such as ConceptNet (Speer et al., 2017), DBpedia (Auer et al., 2007), or domain-specific ontologies. This integration provides a dual-layered benefit: grounding the concepts in authoritative knowledge to reduce hallucinations and enriching sparse textual input with additional context and semantic clarity.

Technically, this approach necessitates a pipeline architecture comprising multiple modules, including an LLM-based concept extractor, an entity linking system that maps extracted terms to entries in a knowledge graph, and a reasoning or enrichment component that expands relationships based on known triples or graph traversals. Semantic similarity measures, often computed using embedding spaces or graph embeddings, are used to match LLM outputs to knowledge base entries with high confidence. Advanced implementations also include disambiguation routines to resolve polysemy in extracted terms. The overall architecture enhances both the precision and explainability of concept maps, making the output more interpretable for end users.

An illustrative example of this approach is the MindMap framework (Wu et al., 2024), which employs ConceptNet to validate and enrich GPT-extracted concepts and their links. This hybrid strategy significantly improved the structural accuracy and pedagogical value of generated maps in classroom use-cases. Additionally, research by Shi et al. (2024) demonstrated that using domain-specific ontologies in combination with LLM outputs could increase concept coverage and reduce noise in concept mapping tasks within biomedical and legal domains.

***Integrated Approach***. This method includes LLMs, rule-based modules, and visualization engines—for the automated generation of concept maps. These frameworks typically segment the pipeline into specialized stages: concept extraction, relationship identification, semantic validation, and graphical rendering. The LLM component, often using models like GPT-4, performs deep contextual analysis to extract candidate concepts and preliminary relationships from raw input text. Subsequently, an intermediate logic module or lightweight symbolic reasoning system evaluates and refines these outputs, applying domain-specific filters or constraints. Another module constructs the graph representation—often in a standardized format like GraphML or JSON-LD—which is then passed to a rendering tool such as Graphviz, D3.js, or Cytoscape for visualization.

The advantage of this architecture lies in its modularity. Each component can be independently upgraded or replaced, allowing the system to evolve as new models or algorithms become available. For example, MapGPT utilizes GPT-based parsing in combination with a graph generation module and a user-friendly web interface that supports real-time editing and feedback. In another implementation, Graphologue, Jiang et al. (2023) merge LLM-extracted entities and relationships with interactive, logic-informed visual diagrams to support exploration and comprehension. These frameworks are particularly suited to educational, clinical, and knowledge management settings, where adaptability, transparency, and user feedback are critical. By decoupling components, modular frameworks also allow researchers to experiment with combinations of LLMs, rule engines, and visualization strategies for domain-specific use cases.

*Linguistic and Heuristic Rule-Based Methods*. Bayrak and Dal (2024) Bayrak and Dal (2024) present a heuristic-based methodology for the automatic generation of concept maps from Turkish texts, addressing the labor-intensive and expertise-dependent nature of manual concept mapping. The approach begins with a preprocessing phase in which sentences are analyzed using the ITU Turkish NLP Pipeline and the Zemberek tool to obtain morphological and syntactic features. Based on dependency parsing, candidate concepts (typically noun or adjective phrases) and their relations (usually verbs) are extracted, with sensitivity to the agglutinative and morphologically rich structure of the Turkish language.

Following concept identification, the algorithm calculates concept frequencies and applies a sentence selection mechanism. Sentences are scored not only by their conceptual richness but also by a reward–penalty system: sentences containing positive relational verbs (e.g., *oluşur*, *ayrılır*) are rewarded, while definitional ones (e.g., *denir*, *tanımlanır*) are penalized. This ensures that the most informative and structurally relevant segments of text contribute to the concept map.

The final step involves sorting selected sentences and converting them into the DOT graph description language, which is then visualized with Graphviz to produce hierarchical maps. The methodology was evaluated on 20 Turkish texts from domains including literature, geography, science, and computer science. Performance was measured with precision, recall, and F-score, showing average concept extraction scores of 0.46 precision, 0.38 recall, and 0.42 F-score, while sentence selection achieved 0.63 precision, 0.53 recall, and 0.54 F-score.

Although results are promising and the automatically generated maps are structurally close to human-authored ones, the approach is limited by the performance of the Turkish NLP tools. In some cases, non-Turkish words are misparsed, or multiple disconnected sub-maps are produced, requiring manual revision. Nonetheless, the method provides a significant step toward scalable, automated concept mapping for Turkish textual analysis.

***Validity evidence of the LLM-generated concept maps***

The validity of LLM-generated concept maps is often assessed through a combination of intrinsic and extrinsic evaluation metrics, each aligned with the technical objectives, computational complexity, intended learning outcomes, and design procedures of the underlying generation methods. These evaluations seek to capture not only the structural and semantic accuracy of the maps but also their pedagogical utility, interpretability, and consistency across varied domains. For fine-tuned LLMs trained on domain-specific datasets, validity is typically measured using a combination of semantic similarity scores, edge accuracy, concept alignment metrics, and structure-based assessments such as graph edit distance or isomorphism measures. These methods assess how closely the generated maps mirror expert-created gold-standard representations and whether they preserve the intended semantic relationships among concepts.

Such models tend to exhibit high internal consistency, contextual coherence, and domain relevance because the training process directly optimizes the model to replicate annotated datasets with expert-defined patterns. Evaluations often also incorporate metrics like node overlap ratio, weighted graph coherence scores, and human-rated alignment indices that reflect both syntactic and semantic fidelity. For instance, Han and Choi (2025) report approximately 83.6 % precision and 74.5 % recall for concept extraction using an LLM-based approach in educational contexts, with topic-agnostic prompts; structural overlap with human-authored maps was not quantified, and no studies yet demonstrate improvements in concept hierarchy depth or connection accuracy through topic-specific fine-tuning. These findings underscore the importance of using rich, task-aligned training data and multi-dimensional validation frameworks to ensure the reliability and relevance of concept maps generated through fine-tuned LLMs.

In contrast, concept maps generated using pre-trained LLMs with prompt engineering are validated primarily through expert judgment or crowd-sourced evaluation, where emphasis is placed on several key aspects such as the breadth of concept coverage, logical coherence of the map, structural soundness of the inferred relationships, and overall alignment with both educational and user-specific expectations. Unlike fine-tuned models that benefit from training on annotated datasets, prompt-based systems are highly sensitive to the construction and clarity of their guiding prompts, meaning even slight variations in phrasing can yield dramatically different outputs.

As a result, validation practices for these systems often incorporate qualitative metrics in addition to basic quantitative indicators. These include expert rubric-based evaluations, user satisfaction surveys, consistency checks across multiple prompt iterations, and cognitive interviews with users to assess how well the generated concept maps support comprehension and learning objectives. Furthermore, evaluation scenarios often involve iterative trials in educational settings, where instructors or students assess the pedagogical relevance of the maps, providing feedback that can be used to fine-tune the prompt templates rather than the model itself.

In a study by ARANGO (2024), for example, LLMapper's outputs were shown to vary significantly in clarity and usefulness depending on the granularity and framing of the prompts. When users were presented with prompts that clearly specified expected output structure—such as indicating the number of layers or explicitly asking for cause-effect relationships—the generated maps were rated significantly higher in usability and conceptual fidelity. These findings underscore the critical role of human-centered prompt design in shaping the success of concept maps derived from pre-trained LLMs and highlight the necessity for iterative evaluation cycles that engage both novice users and domain experts.

For hybrid models that integrate LLMs with external structured knowledge sources, validity evidence encompasses both structural and semantic metrics designed to assess the integrity and utility of the generated concept maps. These models are commonly benchmarked using knowledge graph alignment standards, including node coverage, semantic overlap, grounding fidelity, and relationship accuracy. In particular, evaluations often focus on the model's ability to correctly identify entities and map them to corresponding entries within a structured knowledge base, such as ConceptNet or DBpedia, thereby ensuring that extracted concepts and relationships are not only textually plausible but also semantically sound within the broader domain of discourse. MindMap (Wu et al., 2024), for instance, illustrates the efficacy of this approach, reporting marked increases in pedagogical usability and content coherence when GPT-extracted concepts were validated and enriched through alignment with ConceptNet relations. Additionally, experiments revealed that the incorporation of structured semantic data reduced the incidence of hallucinated concepts and improved users' confidence in map accuracy during instructional tasks.

Integrated modular frameworks, which combine LLMs with rule-based logic engines and dedicated visualization modules, are validated through a multilayered, component-specific process that ensures both functional integrity and holistic usability. Each module within the framework is assessed individually using targeted evaluation protocols—such as measuring the precision and recall of the LLM's concept extraction, verifying the logical validity of relationships inferred by the rule engine, and inspecting the clarity, navigability, and structural layout of the graphical rendering system. This approach allows researchers and developers to isolate errors, optimize subsystems, and iteratively refine each stage of the concept map generation pipeline. Furthermore, end-to-end evaluations often synthesize these outputs into composite validity scores reflecting the pipeline's overall pedagogical effectiveness, semantic completeness, and user acceptability. A representative implementation of this validation paradigm is Graphologue by Jiang et al. (2023), which combines a technical evaluation of GPT-4's inline entity and relationship annotation accuracy (achieving up to 97.2% F-score for entities and 92.4% for relationships after correction) with a user study involving seven experienced ChatGPT users who assessed the usability of diagram-based exploration and comprehension tools . Their results highlight that this layered strategy not only surfaces system-level precision and error types but also demonstrates how synchronized diagrammatic and textual representations improve comprehension, reduce information overload, and support flexible exploration of LLM outputs.

Finally, for linguistic and heuristic rule-based methods such as those of Bayrak and Dal (2023), validation is conducted through precision, recall, and F-score of concept and relation extraction against manually created benchmarks. Their system employs deterministic parsing and morphological analysis in a Turkish NLP pipeline, leveraging the agglutinative structure of the language but limiting portability across domains and languages. Reported results show moderate performance (≈0.42 F-score for concept extraction; ≈0.54 for sentence selection), with occasional manual correction needed for parsing errors and incomplete relations. In contrast to probabilistic and interactive approaches, these rule-based pipelines emphasize transparency and reproducibility, providing interpretable traces of how each concept and relation is derived while facing scalability challenges in adapting to new genres or evolving educational content. Particularly in domain-specific applications, these methods benefit from their interpretability, offering clear traceability of how each concept and relation was derived—an advantage when transparency and auditability are prioritized. However, scalability remains a challenge, particularly in adapting the rule sets to different genres, domains, or evolving educational content without extensive manual re-engineering.

Altogether, these validation practices reflect the diverse nature of LLM-driven concept map generation, with each method necessitating tailored metrics and procedures that correspond to its technical foundation, computational architecture, and degree of user interaction. For example, systems that rely heavily on fine-tuned LLMs require precise performance indicators that can measure model generalization, domain alignment, and the semantic validity of extracted relationships. These might include metrics such as F1-score, precision-recall balance, and graph topology similarity to expert-created maps. On the other hand, methods rooted in prompt engineering prioritize usability, coherence, and the fidelity of structure generation based on minimal guidance—often assessed via qualitative expert review and user-centered scoring rubrics. Hybrid systems incorporating external knowledge graphs necessitate dual-layered validation, measuring both the accuracy of textual concept extraction and the semantic correctness of mapped entities within a knowledge base. For these, grounding fidelity, ontology alignment, and triple validation are common metrics. Modular frameworks must validate each pipeline component individually, demanding custom testbeds and cumulative success indicators across modules like LLM interpretation, rule logic processing, and visual rendering engines. Even rule-based and heuristic approaches are subject to rigorous syntactic and semantic parsing validations, particularly when constrained by language morphology or dependency parsing precision. Ultimately, the evaluation protocols and performance metrics selected must be tightly coupled with each system's architectural priorities, computational strategy, and educational or analytical deployment goals, making validation both a technical and a contextual exercise. and intended application context.

**Discussion**

The analysis of contemporary methods for generating concept maps using LLMs reveals a rapidly diversifying landscape of tools and frameworks, each shaped by distinct technical foundations and pedagogical ambitions. A key finding from this synthesis is the classification of LLM-driven concept map generation methods into six major categories: human-in-the-loop systems, weakly supervised learning models, fine-tuned domain-specific LLMs, pre-trained LLMs with prompt engineering, hybrid systems integrating knowledge bases, and modular frameworks combining symbolic and statistical tools. Each method offers a different balance between automation, interpretability, scalability, and user interaction, reflecting the complex trade-offs in current AI-assisted learning technologies.

These findings contribute to the field by offering a comprehensive taxonomy and comparative analysis of LLM applications in concept mapping, a task critical to knowledge visualization, cognitive scaffolding, and educational assessment. This synthesis aligns with and extends prior work in AI-assisted learning, such as Erdem (2017), who highlighted the need for automated yet interpretable knowledge representation tools, and Shi et al. (2024), who demonstrated the effectiveness of domain-tuned models for biomedical knowledge graphs. The review also builds upon Han and Choi (2025), who presented empirical validation of concept maps generated by fine-tuned models in STEM education, and Chen et al. (2024), who illustrated the pedagogical relevance of few-shot LLM outputs. In contrast to these studies that focused on individual implementations, the present work integrates diverse methods and tools into a comparative framework, thus providing a broader perspective for researchers and practitioners alike.

A major trend in the literature is the increasing integration of LLMs with external tools—such as ontologies, visualization libraries, or symbolic reasoning engines—enabling hybrid and modular systems to bridge the gap between black-box models and interpretable outputs. This trend aligns with broader shifts in the AI and education technology communities toward explainable and human-centered AI. For instance, the MindMap framework (Wu et al., 2024) reflects this movement by using ConceptNet to semantically ground GPT-extracted content, enhancing both reliability and pedagogical coherence. Similarly, tools like Graphologue by Jiang et al. (2023), embody modular validation strategies, a clear advancement from early single-shot LLM outputs. The literature also shows a maturation in evaluation methodologies, shifting from performance-centered benchmarks to layered validation including structural accuracy, semantic depth, and learning utility.

Despite these advances, the review identifies several notable research gaps that remain underexplored in current literature. First, there is a scarcity of longitudinal studies or classroom-based interventions examining how teachers and students use and learn from AI-generated concept maps over time. While studies like Arango (2023) demonstrate usability in controlled settings, they stop short of assessing real-world instructional impact. Second, cross-linguistic adaptability remains a challenge, with most current tools developed for English or morphologically simple languages. The heuristic approach by Bayrak and Dal (2023) represents an important exception, showing that rule-based NLP tailored for Turkish can yield reliable outputs. However, comparable research for other non-English languages remains limited. Third, although prompt engineering has become more nuanced, there is little theoretical grounding or empirical testing on optimal prompt structures across contexts and user types. Finally, while educational relevance is frequently cited, few systems are co-designed with educators or integrated into curriculum-based practices, reflecting a misalignment between technical development and pedagogical application.

Future research should bridge these gaps by incorporating interdisciplinary design approaches, conducting classroom-based evaluations, and prioritizing adaptive scaffolding aligned with learning goals. In particular, research that co-develops concept mapping tools with teachers and learners—evaluated not only on output quality but also on educational outcomes such as knowledge retention, metacognitive skills, and engagement—will be crucial. Furthermore, integrating LLM-generated concept maps into formative assessment systems or adaptive learning platforms could enable personalized feedback and support deeper cognitive engagement. To fully realize these goals, collaboration among AI researchers, education scientists, linguists, and domain experts will be essential. Only through such cross-disciplinary integration can we ensure that advances in LLM technologies are matched by advances in equitable, interpretable, and context-aware learning applications.

**Conclusions**

This review highlights the growing sophistication and diversity of approaches used to generate concept maps with LLMs, offering a comprehensive typology that bridges technical, pedagogical, and domain-specific considerations. The primary takeaway is that LLMs are not monolithic tools but part of a broad ecosystem of technologies that, when combined thoughtfully, can significantly advance the practice of knowledge visualization and cognitive support in educational settings. From weakly supervised learning frameworks and fine-tuned models to prompt-driven generation, hybrid systems, and modular infrastructures, the review illustrates how each method offers unique advantages and constraints depending on the learning context and technological objectives.

Importantly, the findings underscore the field's move toward more explainable, user-centered, and pedagogically aligned AI systems. By integrating external knowledge bases, incorporating human-in-the-loop interfaces, and adopting layered validation strategies, current approaches are aligning more closely with the educational goals of transparency, usability, and contextual adaptability. These developments mark a meaningful departure from earlier opaque or static concept mapping systems, highlighting the potential of LLMs to serve as dynamic cognitive partners rather than mere content extractors.

Despite this progress, significant research challenges remain, particularly around classroom implementation, multilingual applicability, and theoretical grounding for prompt design and user interaction. The review concludes that future work must embrace interdisciplinary collaboration and empirical grounding in educational practice to ensure that advances in LLM capabilities result in meaningful, equitable, and sustainable innovations in teaching and learning. More empirical research is needed to compare the quality, pedagogical utility, and cognitive alignment of LLM-generated concept maps with those produced through traditional, human-centered methods. In addition, research is needed to examine the educational effectiveness of these AI-generated concept maps—such as their capacity to support learning outcomes, foster conceptual understanding, and integrate meaningfully into instructional practices—relative to traditional, manually created maps. As the landscape continues to evolve, the synthesis presented here serves as a foundation for future studies, offering both a roadmap for system development and a call to action for researchers and educators to co-create the next generation of concept map technologies informed by real-world pedagogical needs.